\documentclass{ieeeaccess}
\usepackage{cite}
\usepackage{amsmath,amssymb,amsfonts}
\usepackage{graphicx}
\usepackage{bm}
\usepackage{array}
\usepackage{tabularx}
\newcolumntype{L}[1]{>{\raggedright\arraybackslash}p{#1}}
\newcolumntype{Y}{>{\raggedright\arraybackslash}X}
\makeatletter
\AtBeginDocument{\DeclareMathVersion{bold}
\SetSymbolFont{operators}{bold}{T1}{times}{b}{n}
\SetSymbolFont{NewLetters}{bold}{T1}{times}{b}{it}
\SetMathAlphabet{\mathrm}{bold}{T1}{times}{b}{n}
\SetMathAlphabet{\mathit}{bold}{T1}{times}{b}{it}
\SetMathAlphabet{\mathbf}{bold}{T1}{times}{b}{n}
\SetMathAlphabet{\mathtt}{bold}{OT1}{pcr}{b}{n}
\SetSymbolFont{symbols}{bold}{OMS}{cmsy}{b}{n}
\renewcommand\boldmath{\@nomath\boldmath\mathversion{bold}}}
\makeatother

\begin{document}
\history{Date of publication xxxx 00, 0000, date of current version xxxx 00, 0000.}
\doi{TBD}

\title{A Domain-Aware Pipeline for Structured Extraction and Retrieval-Augmented Question Answering over arXiv Superconductivity Literature}

\author{\uppercase{Ivan Iorsh}\authorrefmark{1},
\uppercase{Aleksandr Kapitonov}\authorrefmark{2},
\uppercase{Ivan Filianin}\authorrefmark{3},
\uppercase{Matvey Lesonen}\authorrefmark{4},
\uppercase{Nikita Chizhov}\authorrefmark{5}}

\address[1]{ITMO University, Saint Petersburg, Russia}
\address[2]{New Uzbekistan University, Tashkent, Uzbekistan (e-mail: a.kapitonov@newuu.uz)}
\address[3]{New Uzbekistan University, Tashkent, Uzbekistan (e-mail: adeptvin1@gmail.com)}
\address[4]{ITMO University, Saint Petersburg, Russia (e-mail: matveylesonen@gmail.com)}
\address[5]{ITMO University, Saint Petersburg, Russia (e-mail: ch.nik.study@gmail.com)}

\markboth
{Iorsh \headeretal: Domain-Aware Pipeline for Superconductivity Literature QA}
{Iorsh \headeretal: Domain-Aware Pipeline for Superconductivity Literature QA}

\corresp{Corresponding authors: Aleksandr Kapitonov (e-mail: a.kapitonov@newuu.uz) and Ivan Iorsh (e-mail: i.iorsh@metalab.ifmo.ru).}

\begin{abstract}
Scientific literature in materials physics is growing faster than researchers can manually inspect it, yet most retrieval tools still rely on keyword matching or generic question answering that ignores domain-specific structure in portable document format files.
This paper presents an end-to-end pipeline that transforms arXiv superconductivity publications into a searchable corpus with structured metadata and retrieval-augmented question answering.
The system combines thematic corpus construction, domain-aware extraction of materials and superconducting transition temperatures, and hybrid retrieval that merges dense chunk embeddings with structured publication fields.
We test whether normalized domain attributes improve evidence retrieval and answer groundedness relative to lexical-only and dense text-only methods.
Extraction is evaluated on a manually verified gold set of 59 superconducting materials, of which 46 have confirmed experimental critical temperatures. On the 33-material primary positive subset, domain-aware pair accuracy reaches 45.5\% versus 9.1\% for a regex baseline.
Retrieval and generation are compared under a fixed language model (\texttt{qwen2.5:7b-instruct}) on a 100-question domain benchmark while varying only the retrieval mechanism: BM25, dense retrieval, hybrid lexical-dense retrieval, and hybrid retrieval with domain metadata.
BM25 gives the highest aggregate source recall (Recall@5 of 47\%), and adding extracted metadata to hybrid retrieval does not improve it (39\% versus 41\%), which we attribute to noise in the extracted fields; the main bottleneck is locating the supporting span inside a correctly retrieved paper.
As a materials-informatics application, literature-derived experimental temperatures are matched to an independent theoretical reference set: after expert alignment, 200 materials appear in both sources, and 20 have a median experimental critical temperature above the theoretical reference value.
\end{abstract}

\begin{keywords}
arXiv, information extraction, retrieval-augmented generation, scientific document processing, superconductivity literature
\end{keywords}

\titlepgskip=-21pt
\maketitle

\section{Introduction}
\label{sec:introduction}

\PARstart{T}{his} paper investigates whether domain-aware structured extraction can improve search quality and grounded answer generation over a specialized corpus of scientific publications. We use arXiv papers on superconductivity as the data source and focus on automatically identifying the studied material, the superconducting critical temperature, system dimensionality, article type, and Debye frequency. These parameters enable analysis of physical properties directly linked to publication content, rather than text matching alone.

The continuous growth of scientific literature makes manual review and cross-paper comparison increasingly inefficient. The problem is especially acute in materials science and condensed-matter physics, where important evidence appears as chemical formulas, abbreviated compound names, numeric properties, and multiple notations for the same material. Traditional search systems rely mainly on titles, abstracts, and word overlap. They therefore miss relevant publications when authors use different wording, chemical symbols, material aliases, or abbreviations.

Scientific portable document format (PDF) files add further difficulty. They contain multi-column layout, tables, mathematical expressions, superscripts and subscripts, line breaks, and chemical formulas. Converting such documents to text may disrupt reading order and distort individual symbols. Technical text extraction alone is insufficient: domain context and links between extracted entities are also required.

This challenge is particularly important for superconductivity literature when analyzing temperature values. A single paper may mention synthesis, annealing, magnetic and structural transition temperatures, Debye temperatures, and results for several materials. The numerically largest temperature in a document is not necessarily the superconducting critical temperature $T_c$. Correct $T_c$ assignment requires analysis of surrounding text and an explicit link between each temperature value and the corresponding material.

Existing search methods address only part of this problem. Keyword search handles paraphrases, abbreviations, and alternative chemical notations poorly. Dense vector retrieval can find semantically similar passages, but textual similarity does not guarantee physical correctness. A retrieved fragment may refer to a different compound, experimental configuration, or type of phase transition. The same limitation appears in retrieval-augmented generation systems. Answer quality depends on which fragments are retrieved and passed to the language model. If the material is misidentified, a chemical formula is corrupted, or a temperature is linked to the wrong physical process, the system may produce a fluent but factually incorrect answer. Generic retrieval-augmented generation approaches usually do not model domain-specific constraints or verify links between materials, properties, and experimental values.

A key research gap is that scientific data extraction, publication search, and answer generation are often treated as separate tasks. Extraction systems produce structured records, whereas search and retrieval-augmented systems work mainly with raw text fragments. Whether normalized domain features can improve document ranking, evidence collection, and answer groundedness remains insufficiently studied.

In this work, we treat these stages as one integrated process. Thematically selected arXiv superconductivity publications are transformed into a corpus in which full text is augmented with structured material and physical-property information. These fields are used together with semantic representations of text fragments. The approach combines dense retrieval with precise domain constraints, such as material name and $T_c$ value.

Our focus is on the information-processing side of the problem: we build the pipeline and then measure, stage by stage, whether domain-oriented structured data helps search and answer generation compared with keyword-only or dense text-only methods.

The main contributions of this paper are as follows:
\begin{enumerate}
\item An end-to-end pipeline covering thematic arXiv selection, scientific document processing, structured extraction, evidence retrieval, and question answering.
\item A domain-aware extraction module for superconducting materials and critical transition temperatures with context filtering to reduce false material--temperature links.
\item A hybrid retrieval approach that combines dense semantic search over scientific text fragments with structured publication metadata, together with an analysis of when such metadata helps and when extraction noise cancels its benefit.
\item An empirical evaluation on a gold set of 59 materials and a 100-question domain benchmark, plus a screening application that compares literature-derived experimental $T_c$ values with an independent theoretical reference set.
\end{enumerate}

The main research question is whether extracted and normalized domain data improve the accuracy of relevant evidence retrieval and answer groundedness compared with lexical-only or dense text-only methods. We separately evaluate the effect of context-aware temperature processing, correct material--$T_c$ pairing, and structured features in chunk ranking.

The remainder of this paper is organized as follows. Section~\ref{sec:related} reviews prior work on scientific information extraction, materials literature processing, scientific search, and retrieval-augmented generation. Section~\ref{sec:overview} presents the overall system architecture. Section~\ref{sec:methods} describes corpus construction, PDF processing, structured extraction, and hybrid retrieval. Section~\ref{sec:experiments} defines the experimental setup. Section~\ref{sec:results} reports evaluation results. Section~\ref{sec:discussion} discusses findings and limitations. Section~\ref{sec:conclusion} concludes the paper.

\section{Related Work}
\label{sec:related}

Prior work spans scientific information extraction, superconductivity-specific corpora, PDF-to-text conversion, document retrieval, and retrieval-augmented generation (RAG). Below we focus on the aspects of each line that matter for a coupled extraction--retrieval pipeline.

\subsection{Structured Information Extraction from Scientific Literature}
\label{sec:related:ie}

ChemDataExtractor~\cite{b1} and ChemDataExtractor~2.0~\cite{b2} established rule-driven pipelines that convert chemical text into database records, including units and property links rather than isolated entity spans. SciBERT~\cite{b3} and MatSciBERT~\cite{b4} later improved materials named-entity recognition and relation classification through domain pretraining. These systems focus on representation and record construction. They do not decide which compound is the primary object of a paper, do not repair PDF conversion artifacts in formulas, and do not test whether extracted fields help downstream retrieval. For our task, rule-based quantity handling and neural semantic matching are complementary.

\subsection{Superconductivity Literature Extraction}
\label{sec:related:sc}

SC-CoMIcs~\cite{b5} showed that specialized annotation is required for superconductor terminology. SuperMat~\cite{b6} further treats critical temperature as a value linked to a material and, when available, to measurement conditions. Grobid-superconductors~\cite{b7} operationalizes that schema over PDF or plain text and powers the large SuperCon2 extraction database, and is the closest extraction analogue to our module. Our contribution is therefore not material--$T_c$ extraction as such, but the use of the extracted fields as retrieval and answering signals, evaluated end to end. The same domain ambiguity noted in these works---identical symbols can denote different physical quantities---motivates the explicit context checks around candidate temperatures used here.

\subsection{Scientific PDF Processing and Corpus Construction}
\label{sec:related:pdf}

Extraction quality is bounded by the initial text representation. Multi-column layout, tables, and broken reading order can split formulas or merge unrelated spans. S2ORC~\cite{b8} demonstrates large-scale conversion of open scientific PDFs into unified full-text corpora. Our setting is narrower: we build a thematic arXiv superconductivity sample so that domain filters and $T_c$ heuristics can be specialized. Domain filtering does not fix layout errors, so formula-preserving PDF conversion remains a separate pipeline stage.

\subsection{Scientific Document Retrieval}
\label{sec:related:retrieval}

Lexical methods such as BM25 remain strong when queries contain exact formulas, but fail under paraphrase. Dense Passage Retrieval~\cite{b9} recovers semantic matches without lexical overlap, yet provides no built-in chemical or physical constraints. Citation-informed encoders such as SPECTER~\cite{b10} and SciNCL~\cite{b11} improve document-level scientific similarity, but are not designed to locate fine-grained experimental claims inside full text. BEIR~\cite{b12} shows that no retriever is universally best across domains. Hybrid sparse--dense retrieval~\cite{b13} is therefore a natural baseline, and we extend it with structured material and $T_c$ metadata as an additional ranking signal.

\subsection{Retrieval-Augmented Generation for Science}
\label{sec:related:rag}

Lewis \emph{et al.}~\cite{b14} established RAG as parametric generation conditioned on a non-parametric index. Correctness still breaks down when retrieval returns a thematically related but physically wrong span. QASPER~\cite{b15} evaluates supporting evidence as well as final answers, and PaperQA~\cite{b16} applies RAG to full-text scientific papers with source attribution. These systems handle domain structure implicitly through text and do not expose normalized superconductivity fields as filters or reranking features.

\subsection{Research Gap}
\label{sec:related:gap}

Extraction systems typically end at databases, whereas scientific RAG systems typically start from generic full-text chunks. The effect of normalized material--$T_c$ fields on evidence ranking and grounded answering has therefore received little measurement. We evaluate extraction, hybrid retrieval with and without domain metadata, and answer groundedness in one pipeline. The hypothesis is that such fields reduce physically irrelevant evidence relative to keyword-only or dense text-only retrieval, provided the extractor is accurate enough.

\begin{table}[t]
\caption{Comparison of representative prior work and the proposed pipeline}
\label{tab:related}
\centering
\renewcommand{\arraystretch}{1.2}
\setlength{\tabcolsep}{3pt}
\begin{tabular}{|l|c|c|c|c|}
\hline
\textbf{Work} & \textbf{IE} & \textbf{Retr.} & \textbf{RAG} & \textbf{Meta.} \\
\hline
ChemDataExtractor~\cite{b1} & \checkmark & -- & -- & -- \\
Grobid-SC~\cite{b7} & \checkmark & -- & -- & -- \\
SPECTER / DPR~\cite{b9,b10} & -- & \checkmark & -- & -- \\
Sparse+dense~\cite{b13} & -- & \checkmark & -- & text \\
PaperQA~\cite{b16} & -- & \checkmark & \checkmark & -- \\
\textbf{This work} & \checkmark & \checkmark & \checkmark & fields \\
\hline
\end{tabular}
\vspace{0.35em}
\begin{minipage}{\columnwidth}
\footnotesize
IE: information extraction.
Retr.: retrieval.
Meta.: structured metadata used in retrieval
(``text'' means sparse+dense text only;
``fields'' means sparse+dense retrieval plus extracted material and $T_c$ metadata).
\end{minipage}
\end{table}

\section{System Overview}
\label{sec:overview}

The proposed system transforms thematically selected arXiv superconductivity publications into a searchable corpus with structured metadata and retrieval-augmented question answering. Inputs are publication records containing an arXiv identifier, title, and topical categories, together with the corresponding PDF files stored in object storage. The system supports both full-prefix processing of all documents under a given storage path and selective reprocessing of individual arXiv identifiers, which avoids restarting the entire pipeline when only a subset of papers must be updated.

The pipeline consists of four linked stages. First, a thematic superconductivity corpus is constructed from arXiv metadata and bulk PDF archives without downloading the full arXiv collection. Second, each selected PDF file is converted to text and processed by domain-aware extractors that recover material, critical temperature $T_c$, article type, dimensionality, unconventional-superconductivity cues, and Debye frequency. Third, bibliographic fields and normalized subject attributes are stored in PostgreSQL, with subject attributes kept as JSONB payloads linked to the source document. Fourth, publications with reliable material extraction are chunked, embedded, and indexed for hybrid retrieval: dense similarity over text fragments is combined with structured metadata filtering and domain-aware reranking, after which a local language model generates answers grounded only in the retrieved context and source identifiers.

\begin{figure}[t!]
  \centering
  \includegraphics[width=\linewidth]{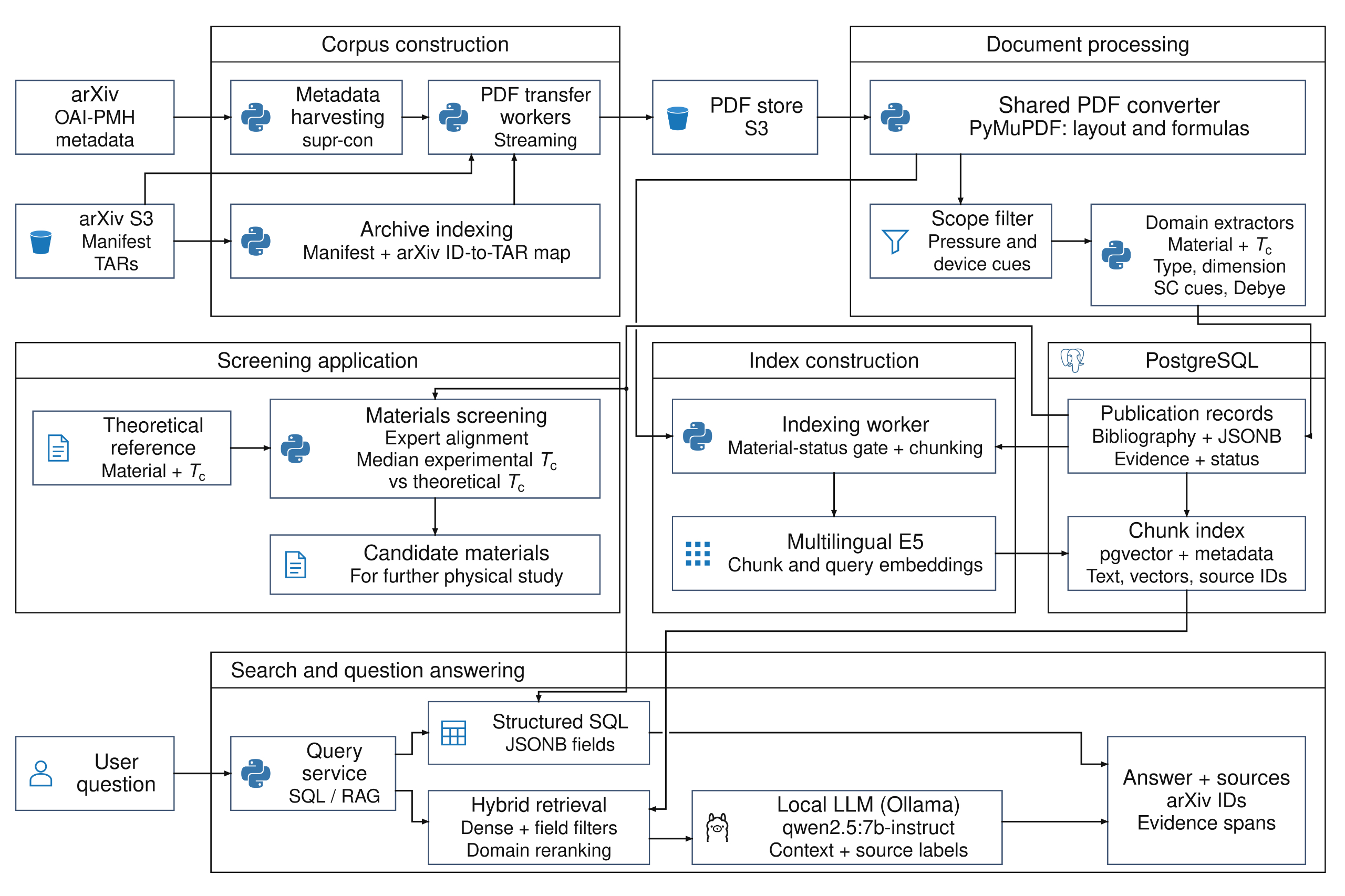}
  \caption{End-to-end architecture of the proposed pipeline, from thematic arXiv corpus construction to hybrid retrieval and grounded answer generation.}
  \label{fig:architecture}
\end{figure}

Extraction and retrieval are deliberately coupled. Extraction results determine which publications enter the main retrieval-augmented generation index. Extracted attributes are copied into chunk metadata for filtering and reranking. A shared PDF converter keeps chemical formulas and reading order consistent across extraction and indexing. The core implementation uses Python, PostgreSQL with pgvector, object storage for PDF files, and a local language model (\texttt{qwen2.5:7b-instruct}) served through Ollama. Fig.~\ref{fig:architecture} shows the end-to-end data flow of the pipeline.

\begin{table}[t]
\caption{Pipeline stages, inputs, and outputs}
\label{tab:stages}
\centering
\setlength{\tabcolsep}{3pt}
\begin{tabularx}{\columnwidth}{|L{0.22\columnwidth}|Y|Y|}
\hline
\textbf{Stage} & \textbf{Input} & \textbf{Output} \\
\hline
Corpus construction & arXiv OAI-PMH metadata, bulk TAR archives & Thematic PDF set in object storage \\
Extraction & PDF + arXiv identifier & Full text + JSONB subject payload \\
Storage & Bibliographic fields + payload & PostgreSQL publication record \\
Indexing and RAG & Text + payload & Chunk embeddings, hybrid retrieval, grounded answer + sources \\
\hline
\end{tabularx}
\end{table}

\section{Methods}
\label{sec:methods}

\subsection{Corpus Construction}
\label{sec:methods:corpus}

The goal of corpus construction is to build a reproducible specialized set of arXiv superconductivity publications without fully downloading and processing the entire arXiv archive. The resulting corpus is the input to document processing, structured information extraction, and retrieval-augmented question answering.

The first stage performs thematic selection from arXiv metadata harvested through the Open Archives Initiative Protocol for Metadata Harvesting (OAI-PMH). Harvesting uses the resumption-token mechanism, which makes it possible to process large metadata volumes sequentially without loading the full archive at once. After metadata retrieval, publications are filtered by the category \texttt{cond-mat.supr-con}. For each retained paper, a normalized arXiv identifier and the bibliographic fields required for later processing are stored.

Individual PDF downloads from arXiv were not used, because the complete PDF collection is distributed as TAR archives. Instead, a two-stage indexing mechanism locates target documents inside the bulk archive. First, the arXiv manifest file is processed to describe the available TAR archives and their parameters. Second, each TAR archive is scanned and an index is built that maps every PDF identifier to its location inside a specific TAR file. This index resolves the physical location of a required document without scanning all archives for every request.

After the index is available, selective extraction is performed. For each paper retained by metadata filtering, the system identifies the corresponding TAR archive and extracts only the needed PDF file into the project object store. Compared with downloading the full arXiv PDF collection, this approach substantially reduces storage and transfer costs by retaining only target-domain publications.

The corpus construction pipeline therefore consists of four steps: (i)~OAI-PMH metadata harvesting, (ii)~filtering by \texttt{cond-mat.supr-con}, (iii)~construction of a complete arXiv-identifier-to-TAR-location index, and (iv)~selective extraction of required PDF files into project storage.

The resulting index contains 2\,921\,746 PDF identifiers distributed across 10\,847 TAR archives. Index correctness was checked by comparing the number of indexed records with the source arXiv manifest. After category filtering, 46\,746 superconductivity-related publications were obtained. Of these, 46\,177 PDF documents were successfully extracted from the bulk store and saved for further processing. The remaining 569 publications appear in arXiv metadata, but the corresponding PDF files were absent from the bulk store at download time. Additional verification confirmed that these files are missing from the source store and are therefore not pipeline extraction failures.

Long-running extraction jobs are managed through a task-state model. Each document-extraction operation stores its current status, which enables parallel processing by multiple workers and safe continuation after transient failures. The design is intended to scale corpus construction without tying the method to a specific cloud vendor.

\subsection{Thematic Filtering of Candidate Publications}
\label{sec:methods:filter}

After a PDF file is converted to text, an additional subject filter is applied. In the current superconductivity block, the beginning of each document is inspected for terms that indicate a mismatch with the target research scope, including mentions of qubits, Josephson devices, SQUID systems, and Majorana states. Papers that contain explicit pressure values in GPa or kbar are also excluded from the current analysis set, because high-pressure experiments introduce temperature and material behavior that fall outside the present extraction rules.

Each document receives one of three outcomes: successful passage, exclusion by the subject filter, or processing error. The output of this stage is therefore a set of arXiv identifiers linked to publication metadata and to the location of the corresponding PDF files in object storage.

\subsection{PDF Processing and Domain-Aware Extraction}
\label{sec:methods:extraction}

For each selected publication, the PDF file is loaded from object storage and converted to text with a PyMuPDF-based extractor. Processing operates on text blocks, lines, and spans rather than as a naive character stream. The algorithm accounts for multi-column layout, hyphenated line breaks, and superscripts and subscripts. For two-column documents, reading order among text blocks is restored, hyphenated words are rejoined, and indices are preserved in the text representation through lightweight HTML markup. Preserving subscripts and superscripts is essential for chemical formulas.

Warnings and critical errors raised during PDF processing are logged separately for each arXiv identifier. Failure on one document does not stop the pipeline for the remaining publications.

The recovered text is passed to a set of domain-oriented extractors that determine:
\begin{itemize}
\item the studied superconducting material or material family,
\item the superconducting critical temperature $T_c$,
\item the article type,
\item the dimensionality of the studied system,
\item cues of unconventional superconductivity,
\item Debye frequency values reported in the paper.
\end{itemize}

Material and $T_c$ extraction are performed jointly in a coupled sequence. First, candidate material mentions are collected and ranked. Next, a preliminary $T_c$ value is extracted. The main material choice is then refined using the temperature evidence, after which $T_c$ is extracted again for the selected material. Together with the final numeric $T_c$ value, the system stores the supporting sentence, surrounding context, evidence type, and a confidence score. These diagnostics support later verification and error analysis.

The output of this stage is the full publication text and a normalized subject payload containing $T_c$, article type, material, dimensionality, an unconventional-superconductivity flag, Debye frequency, material-determination status, and diagnostic information about $T_c$ extraction.

\begin{table}[t]
\caption{Extracted subject fields stored in the publication payload}
\label{tab:fields}
\centering
\setlength{\tabcolsep}{3pt}
\begin{tabularx}{\columnwidth}{|L{0.28\columnwidth}|Y|}
\hline
\textbf{Field} & \textbf{Description} \\
\hline
Material & Formula, alias, or material family, with determination status \\
$T_c$ & Critical temperature in kelvin, with supporting context and confidence \\
Article type & Experimental, theoretical, or mixed \\
Dimensionality & Bulk, two-dimensional / thin-film, or mixed \\
Unconventional SC & Presence of unconventional-superconductivity cues \\
Debye frequency & Reported Debye parameter with unit, kept separate from $T_c$ \\
\hline
\end{tabularx}
\end{table}

\subsection{Structured Storage}
\label{sec:methods:storage}

The system separates storage of source documents, article-level attributes, and search representations. Source PDF files remain in object storage, while publication metadata, processing statuses, and subject-extraction results are stored in PostgreSQL. Stable bibliographic fields, such as the arXiv identifier and bibliographic descriptors, are stored as relational columns. Subject attributes are stored as a JSONB payload, which allows the set of extracted characteristics to evolve without creating a separate table for each new field.

Before persistence, values are normalized: missing and invalid fields are mapped to a predictable schema, illegal control characters are removed, and material extraction receives an explicit status. Subject-extraction status and search-index status are stored separately. The result is a structured publication record keyed by arXiv identifier and containing bibliographic data, a normalized JSONB subject payload, processing status, and a link to the source PDF file.

\subsection{Chunking and Vector Index Construction}
\label{sec:methods:index}

Only publications for which material extraction finished with an admissible status are included in the vector index. Their PDF files are converted again with the same text extractor used for subject extraction, which keeps chemical formulas and reading order consistent across pipeline stages.

The recovered text is cleaned and split into overlapping fragments. In the current configuration, each fragment is approximately 1600 characters long, with a 250-character overlap between neighbors. When suitable boundaries exist, the split point is moved to the end of a paragraph, sentence, or other logically complete unit.

Each fragment is embedded with a multilingual E5-family model. Vectors are stored in PostgreSQL using the pgvector extension. A fragment record contains the arXiv identifier, fragment ordinal, source text, embedding, embedding-model name, text-extractor version, and a JSONB metadata object. The metadata copies the publication title, categories, recovered materials, $T_c$, dimensionality, article type, and unconventional-superconductivity flag. Thus every text fragment remains linked both to its source publication and to that publication's structured attributes.

\subsection{Hybrid Retrieval and Answer Generation}
\label{sec:methods:rag}

User queries are handled in two complementary modes. Explicit structured requests, such as finding materials with $T_c$ above a given threshold or selecting publications of a given dimensionality, are answered with deterministic SQL over JSONB fields. In that case, the answer is produced directly from previously extracted structured data.

For open scientific questions, semantic search over text fragments is used. The query is embedded with the same model used during indexing. Domain constraints are also inferred from the query text, including material, $T_c$ range, dimensionality, article type, and unconventional-superconductivity cues. These constraints are converted into metadata filters over fragment records.

After the initial vector search, results are reranked. Fragments that consist mainly of bibliographies, figure captions, tables, corrupted text, or dense formulas without coherent explanation are down-weighted. Fragments that explicitly mention the requested material or confirm a critical-temperature value are up-weighted. To improve source diversity, the number of fragments selected from a single publication is limited. If strict metadata filtering returns too few candidates, vector search is repeated without mandatory constraints, and the same domain-aware reranker is applied.

Selected fragments are grouped by publication and converted into language-model context. Each paper receives a source label such as \texttt{S1} or \texttt{S2}. Structured article attributes and the strongest supporting spans are passed together with the fragment text. A local language model (\texttt{qwen2.5:7b-instruct}) receives the user question and the assembled context, with an instruction not to invent materials, temperature values, or conclusions that are absent from the provided sources. The system returns a textual answer together with the used publications and the identifiers of the fragments that support the answer.

\subsection{Coupling of Extraction and Retrieval}
\label{sec:methods:coupling}

Information extraction and search are designed as interdependent parts of one pipeline. First, subject-extraction results determine membership in the main retrieval-augmented generation index: publications without a reliably determined material are not indexed. Second, extracted attributes are copied into the metadata of every text fragment, which enables SQL filtering, deterministic answers, and domain-aware reranking of semantic search. Final ranking therefore depends not only on vector similarity between the query and publication text, but also on subject attributes recovered earlier. Third, PDF conversion quality affects both subsystems: loss of chemical subscripts can corrupt material extraction, and broken column order can produce incoherent search chunks. Using one domain-aware PDF converter keeps textual representation consistent across extraction and indexing.

\section{Experimental Setup}
\label{sec:experiments}

This section specifies the evaluation protocol: the corpus, gold-set annotation, extraction scoring, retrieval and generation setups, baselines, and ablations. Numerical results are reported in Section~\ref{sec:results}.

\subsection{Corpus}
\label{sec:experiments:dataset}

The evaluation uses the arXiv superconductivity corpus constructed as described in Section~\ref{sec:methods:corpus}; its statistics are summarized in Table~\ref{tab:corpus}. The 569 metadata records without a PDF file in the arXiv bulk store were excluded from document-level processing.

\begin{table}[t]
\caption{Corpus statistics}
\label{tab:corpus}
\centering
\setlength{\tabcolsep}{4pt}
\begin{tabularx}{\columnwidth}{|Y|r|}
\hline
\textbf{Quantity} & \textbf{Value} \\
\hline
Indexed PDF identifiers in arXiv bulk store & 2\,921\,746 \\
TAR archives indexed & 10\,847 \\
Publications in category \texttt{cond-mat.supr-con} & 46\,746 \\
PDF files successfully extracted & 46\,177 \\
Missing PDF files in bulk store & 569 \\
PDF coverage of the thematic set & $98.8\%$ \\
\hline
\end{tabularx}
\end{table}

\subsection{Gold Set for Extraction Evaluation}
\label{sec:experiments:gold}

Extraction quality is measured on a manually verified gold set of 59 superconducting materials. The unit of evaluation is a material record, not an arbitrary paper sample. Each record is used to check three decisions: identification of the studied material, extraction of superconducting $T_c$, and correctness of the material--$T_c$ link.

For every record we store a normalized material designation, chemical formula, aliases when present, $T_c$ when available, an arXiv publication that contains the corresponding experimental evidence, and the outcome of manual inspection of automatic extraction. Experimental $T_c$ was confirmed for 46 of the 59 materials. For 33 materials, an arXiv publication illustrating correct automatic extraction was also recorded. Manually inspected incorrect extractions were retained for error analysis.

\begin{table}[t]
\caption{Gold-set statistics for extraction evaluation}
\label{tab:gold}
\centering
\setlength{\tabcolsep}{4pt}
\begin{tabularx}{\columnwidth}{|Y|r|}
\hline
\textbf{Quantity} & \textbf{Value} \\
\hline
Manually verified materials & 59 \\
Materials with confirmed experimental $T_c$ & 46 \\
Materials with an arXiv example of correct extraction & 33 \\
\hline
\end{tabularx}
\end{table}

The primary material is the direct object of study in the source publication. Compounds mentioned only for comparison, literature review, or historical context are not treated as the main material. Only superconducting transition temperature is accepted as $T_c$. Debye temperatures, structural and magnetic transitions, and synthesis, annealing, or other processing temperatures are excluded. A numerically correct temperature is not sufficient: the value must belong to the annotated material and must describe the superconducting transition. Formulas, abbreviations, and aliases are normalized before matching and are treated as equivalent only when the publication context makes the correspondence unambiguous.

The scoring record stores the fields
\texttt{arxiv\_id}, \texttt{material\_gold}, \texttt{tc\_gold},
\texttt{link\_ok}, and \texttt{notes}.

\subsection{Extraction Evaluation Protocol}
\label{sec:experiments:extraction-eval}

The full domain-aware extractor is scored at three independent levels. Material identity is correct after formula and alias normalization when the prediction matches the gold material. $T_c$ is correct when
\begin{equation}
\left|T_c^{\mathrm{pred}}-T_c^{\mathrm{gold}}\right| \leq \Delta T,
\end{equation}
where $\Delta T=8\,\mathrm{K}$ is the pre-established tolerance for rounding and reporting variation. Material--$T_c$ linkage is correct only when the recovered temperature both matches the gold material and refers to the superconducting transition.

Evaluation is reported on two complementary subsets of the gold set. The primary positive set contains the 33 materials with an expert-linked publication for which a correct extraction is expected. On this set we report material accuracy, $T_c$ accuracy, and pair accuracy. The hard-case set contains 22 expert-labeled failure cases and is used to measure rejection of incorrect temperatures and to classify remaining errors into three types. Type~1 errors assign a non-superconducting transition temperature (for example, a magnetic or structural transition) to the correctly identified material. Type~2 errors attach a superconducting $T_c$ of a different compound mentioned in the same paper to the extracted formula. Type~3 errors misclassify the article type, for example labeling a theoretical paper as experimental or hybrid.

The full extractor is compared with three simpler configurations. Regex-based extraction searches for formula-like strings and temperature patterns without contextual validation. Extraction without filters disables heuristic temperature and context checks. Extraction without material--$T_c$ binding disables mutual refinement of material and temperature while retaining the remaining filters.

Manual error analysis records false $T_c$, incorrect main-material choice, corrupted or incorrectly normalized formulas, missed $T_c$, and incorrect material--$T_c$ association.

\subsection{Question Answering Benchmark}
\label{sec:experiments:qa}

Retrieval and retrieval-augmented generation are evaluated on a domain-specific question-answering benchmark of 100 questions built from indexed corpus publications. Each question stores a question identifier, the question text, a reference answer, one or more supporting arXiv identifiers, and one to three supporting evidence spans. Questions without direct textual evidence in the corpus were excluded from evaluation. This representation separates three evaluation levels: finding the correct publication, finding the supporting span, and generating a grounded answer.

The benchmark composition is summarized in Table~\ref{tab:qa-types}.

\begin{table}[t]
\caption{Composition of the 100-question QA benchmark}
\label{tab:qa-types}
\centering
\setlength{\tabcolsep}{4pt}
\begin{tabular}{|l|r|}
\hline
\textbf{Question type} & \textbf{Count} \\
\hline
$T_c$ lookup & 33 \\
Material-property lookup & 20 \\
Source grounding & 18 \\
Threshold / group & 14 \\
Comparison & 10 \\
Open evidence & 5 \\
\hline
Total & 100 \\
\hline
\end{tabular}
\end{table}

\subsection{Retrieval and RAG Evaluation}
\label{sec:experiments:retrieval}

Four retrieval configurations are compared: BM25, dense retrieval, hybrid lexical and dense retrieval without subject metadata~\cite{b13}, and hybrid retrieval with domain metadata (\texttt{hybrid\_fields}). All configurations use the chunking described in Section~\ref{sec:methods:index}.

Source retrieval is reported with Recall@$1$, Recall@$5$, and mean reciprocal rank (MRR). Evidence quality is reported with Evidence Recall@$10$ (Ev@$10$), which requires hitting a manually annotated supporting span rather than only the correct arXiv source.

For generation, all RAG experiments used the same local language model served through Ollama, namely \texttt{qwen2.5:7b-instruct}. Generation parameters were fixed across retrieval configurations: temperature~$0.0$, context length~$8192$, $\mathrm{top}\text{-}p=1.0$, and at most~$512$ generated tokens. The same prompt template was used in every run. Only the retrieval mechanism changed between configurations. The model was instructed to answer only from the retrieved context and to abstain when the evidence was insufficient. Answer accuracy measures agreement with the reference answer over all questions. Groundedness is computed only among non-abstaining answers and measures whether factual claims are supported by retrieved evidence~\cite{b15}. Abstention rate records cases in which the model declines to answer. We additionally report answer accuracy conditioned on gold evidence being present in the final language-model context, which isolates generation quality from retrieval failures. Two successful examples and one failure case are traced through the full cascade
\begin{equation*}
\begin{aligned}
&\mathrm{PDF}\!\to\!\mathrm{text}\!\to\!\mathrm{extraction}\!\to\!\mathrm{metadata}\\
&\qquad\to\mathrm{chunks}\!\to\!\mathrm{retrieval}\!\to\!\mathrm{evidence}\!\to\!\mathrm{answer},
\end{aligned}
\end{equation*}
so that a generation error can be distinguished from a failure at an earlier pipeline stage.

\section{Results}
\label{sec:results}

\subsection{Corpus and Processing}
\label{sec:results:corpus}

As shown in Table~\ref{tab:corpus}, PDF files were obtained for $98.8\%$ of the thematically selected publications; all 569 missing files are absent from the arXiv bulk store itself.

\subsection{Extraction Results}
\label{sec:results:extraction}

Extraction is reported separately for material identification, $T_c$ extraction, and material--$T_c$ linkage. Table~\ref{tab:extraction} summarizes accuracy on the primary positive set of 33 expert-linked materials. Domain-aware extraction substantially outperforms regex: pair accuracy rises from $9.1\%$ to $45.5\%$. Material accuracy reaches $81.8\%$ for all non-regex configurations, so further gains must come from temperature selection rather than material recognition.

\begin{table}[t]
\caption{Extraction accuracy on the primary positive set of 33 expert-linked materials.}
\label{tab:extraction}
\centering
\setlength{\tabcolsep}{3pt}
\begin{tabular}{|L{0.46\columnwidth}|c|c|c|}
\hline
\textbf{System} & \textbf{Material} & \textbf{T\textsubscript{c}} & \textbf{Pair} \\
\hline
Regex-based extraction & $24.2\%$ & $36.4\%$ & $9.1\%$ \\
Without filters & $81.8\%$ & $60.6\%$ & $54.5\%$ \\
Without material--$T_c$ binding & $81.8\%$ & $51.5\%$ & $42.4\%$ \\
Full domain-aware extractor & $81.8\%$ & $54.5\%$ & $45.5\%$ \\
\hline
\end{tabular}
\end{table}

Removing filters yields higher $T_c$ and pair accuracy than the full pipeline ($60.6\%$/$54.5\%$ versus $54.5\%$/$45.5\%$). Inspection of disagreements shows that final heuristics sometimes over-reject correct candidates or shift the selected value. For NbS$_2$ (gold $6.2\,\mathrm{K}$), the value is recovered without filters but suppressed (null) under the full system. For NiS$_2$ (gold $1.5\,\mathrm{K}$) and YPd$_2$Sn (gold $5.4\,\mathrm{K}$), the full system selects a different reported value ($1.8\,\mathrm{K}$ and $4.0\,\mathrm{K}$, respectively). The opposite also occurs: for RuB$_2$, the unfiltered extractor returns $20\,\mathrm{K}$ while the full system recovers the gold value $1.5\,\mathrm{K}$. Material--$T_c$ binding gives a modest gain over the no-binding ablation ($45.5\%$ versus $42.4\%$ pair accuracy). For IrTe$_2$, binding recovers the gold value $4.48\,\mathrm{K}$ that is otherwise missed. Overall, filtering can suppress false temperatures, but the present heuristics trade recall for precision and occasionally select incorrect candidates.

Table~\ref{tab:extraction-hard} reports hard-case behavior on 22 expert-labeled failures. Bad-$T_c$ rejection counts a case as rejected when the extractor does not return the specifically incorrect temperature. Both a corrected value and a null prediction count as rejection. Regex therefore appears strong ($87.5\%$ rejection), mainly because it often returns nothing. Columns T1--T3 count how many cases of each error class are avoided or corrected under each method, using the taxonomy in Section~\ref{sec:experiments:extraction-eval}.

\begin{table}[t]
\caption{Hard-case set of 22 expert-labeled failures. Bad-$T_c$ rejection treats null and corrected predictions as successful rejection of the annotated wrong temperature.}
\label{tab:extraction-hard}
\centering
\setlength{\tabcolsep}{2pt}
\small
\begin{tabularx}{\columnwidth}{|Y|c|c|c|c|}
\hline
\textbf{System} & \textbf{Bad-$T_c$} & \textbf{T1} & \textbf{T2} & \textbf{T3} \\
\hline
Regex-based extraction & $87.5\%$ & $7/9$ & $7/7$ & $1/5$ \\
Without filters & $50.0\%$ & $5/9$ & $3/7$ & $1/5$ \\
Without material--$T_c$ binding & $56.2\%$ & $5/9$ & $4/7$ & $1/5$ \\
Full domain-aware extractor & $50.0\%$ & $5/9$ & $3/7$ & $1/5$ \\
\hline
\multicolumn{5}{|p{0.96\columnwidth}|}{\footnotesize T1: non-superconducting transition for the correct material. T2: superconducting $T_c$ of another material. T3: wrong article type.} \\
\hline
\end{tabularx}
\end{table}

\subsection{Retrieval and Question Answering Results}
\label{sec:results:qa}

Table~\ref{tab:retrieval} compares the four retrieval configurations on the 100-question benchmark. No single method dominates all metrics. BM25 obtains the strongest overall source retrieval (Recall@$5=47.0\%$, MRR~$=0.366$), largely because 33 questions are direct $T_c$ lookups for a named material, where the chemical formula is a strong lexical marker (BM25 Recall@$5=66.7\%$ on that subset). Dense retrieval is weaker overall but better on open-evidence questions (Recall@$5=40.0\%$ versus $0.0\%$ for BM25), although this subset contains only five questions. Hybrid retrieval helps on multi-document comparison and threshold queries. Comparison Recall@$5$ rises from $20.0\%$ for BM25 and dense to $30.0\%$ for hybrid. Threshold Recall@$5$ rises from $7.1\%$ (BM25) and $21.4\%$ (dense) to $28.6\%$ (hybrid).

\begin{table*}[t]
\caption{Retrieval and RAG comparison on the 100-question domain benchmark. Groundedness is computed only over non-abstaining answers.}
\label{tab:retrieval}
\centering
\setlength{\tabcolsep}{6pt}
\begin{tabular}{|l|c|c|c|c|c|c|c|}
\hline
\textbf{System} & \textbf{R@$1$} & \textbf{R@$5$} & \textbf{MRR} & \textbf{Ev@$10$} & \textbf{Ans.} & \textbf{Grnd.} & \textbf{Abs.} \\
\hline
BM25 & $28.0\%$ & $47.0\%$ & $0.366$ & $23.0\%$ & $4.0\%$ & $83.3\%$ & $88.0\%$ \\
Dense & $23.0\%$ & $35.0\%$ & $0.333$ & $23.0\%$ & $2.0\%$ & $100\%$ & $92.0\%$ \\
Hybrid & $23.0\%$ & $41.0\%$ & $0.332$ & $14.0\%$ & $2.0\%$ & $100\%$ & $92.0\%$ \\
Hybrid+fields & $24.0\%$ & $39.0\%$ & $0.319$ & $14.0\%$ & $4.0\%$ & $100\%$ & $84.0\%$ \\
\hline
\end{tabular}
\end{table*}

Hybrid retrieval with extracted material/$T_c$ fields does not improve on plain hybrid overall (Recall@$5=39\%$ versus $41\%$). This is consistent with the extraction results: imperfect material and $T_c$ metadata can demote the correct paper as easily as they promote it. Answer accuracy remains low because the generator is conservative and abstains when evidence is insufficient. For hybrid with fields, all 16 non-abstaining answers are grounded in the retrieved context, but only 4 match the reference answer; grounded but incorrect answers arise when the retrieved span is genuine but refers to a different sample, compound, or transition. In a scientific setting, abstention is preferable to unsupported answers, even though it reduces the fraction of fully solved questions.

When gold evidence is already present in the language-model context, answer accuracy rises. On these subsets, hybrid with fields reaches $33.3\%$, versus $18.2\%$ for BM25, $16.7\%$ for hybrid, and $12.5\%$ for dense; because the subsets are small, this difference is indicative rather than conclusive. The main bottleneck is evidence selection rather than generation: maximum source Recall@$5$ is $47\%$, whereas maximum Evidence Recall@$10$ is only $23\%$, so finding the correct publication is substantially easier than locating the exact supporting span within it.

\subsection{Qualitative Answer Cases}
\label{sec:results:cases}

Two successful cases show that the full cascade can operate end to end. For Li$_2$IrSi$_3$, the gold $T_c$ is $3.7\,\mathrm{K}$. Hybrid retrieval with material/$T_c$ fields ranked arXiv~1409.3662 first, placed the supporting span in the model context, and produced the correct answer $3.7\,\mathrm{K}$ with a source citation. For ZrNi$_2$Ga, the gold $T_c$ is $2.9\,\mathrm{K}$. BM25 ranked arXiv~0808.2356 first, passed the supporting span to the model, and recovered $2.9\,\mathrm{K}$.

A failure on NbB$_2$ isolates the evidence-ranking problem. The gold $T_c$ is $9.5\,\mathrm{K}$ from arXiv~0811.4224. Dense retrieval placed the correct paper second, so document-level search had not failed. Evidence Recall@$10$ was nevertheless zero for this question: the required span never entered the model context. The model instead received a fragment containing $39.0\,\mathrm{K}$ and answered that NbB$_2$ has $T_c=39.0\,\mathrm{K}$. The error is therefore neither an unconstrained hallucination nor a document miss, but a wrong intra-document evidence choice that propagates into a fluent but incorrect answer.

\section{Discussion}
\label{sec:discussion}

Domain-aware processing is especially important for superconductivity literature because a single publication may mention several materials and many temperatures that belong to different physical processes. Contextual $T_c$ filtering is intended to exclude Debye, structural, magnetic, synthesis, and other non-superconducting temperatures. Material--$T_c$ binding addresses the additional problem of attaching a recovered value to the correct studied compound. The extraction experiments confirm a large gain over regex on material--$T_c$ pairs, but also show that the current final filters can over-reject correct candidates. Improving the precision--recall balance of these heuristics is therefore a priority for the extractor.

Structured metadata also makes downstream retrieval depend on extraction quality. Correctly extracted material and $T_c$ fields can improve filtering and ranking of evidence, whereas an extraction error can be amplified later along the cascade defined in Section~\ref{sec:experiments:retrieval}. Empirically, hybrid retrieval with extracted fields did not outperform plain hybrid on aggregate source recall, which is consistent with noisy metadata. When gold evidence did reach the generator, hybrid with fields gave the highest conditional answer accuracy, although on small subsets. The NbB$_2$ failure further shows that document-level recall is not enough: incorrect intra-document evidence can still yield a grounded-looking but wrong temperature.

This is the main difference from specialized extraction systems such as Grobid-superconductors~\cite{b7}. The present work does not stop at constructing a database of extracted materials and $T_c$ values: those structured fields are used as signals for evidence retrieval and grounded question answering, and their value is limited by extraction fidelity.

\subsection{Application: Screening of Anomalous Superconductivity Candidates}
\label{sec:discussion:anomalies}

As a separate materials-informatics application, the pipeline was used to find materials whose literature-derived experimental $T_c$ disagrees with theoretical reference values. Two independent datasets were compared.

Dataset~1 is a theoretical reference set of studied superconducting materials. For each compound it stores chemical composition and a theoretically computed critical temperature $T_c^{\mathrm{theory}}$. Dataset~2 is a literature-derived experimental set produced automatically by the proposed pipeline. For each material the system collects experimental $T_c$ values found across publications, together with their sources.

A physics expert then aligned the two sets. Materials were first reduced to a comparable representation using chemical formulas and aliases. The expert then verified correspondences between the theoretical reference set and the literature-extraction results. When a material appeared in more than one publication, the median experimental critical temperature was used. For observations $T_{c,1}^{\mathrm{exp}},\ldots,T_{c,n}^{\mathrm{exp}}$,
\begin{equation}
\widetilde{T}_c^{\mathrm{exp}}
=\operatorname{median}\bigl(T_{c,1}^{\mathrm{exp}},\ldots,T_{c,n}^{\mathrm{exp}}\bigr).
\end{equation}
The median reduces the influence of outliers, incorrectly extracted values, and sample-to-sample experimental differences.

After this expert matching, 200 materials were present in both datasets. For each of them the median experimental value was compared with the corresponding theoretical value. A material was treated as a potentially anomalous candidate if
\begin{equation}
\widetilde{T}_c^{\mathrm{exp}} > T_c^{\mathrm{theory}}.
\end{equation}
Of the 200 matched materials, 20 had a median literature $T_c$ above the theoretical reference. These 20 compounds were retained as candidates for further investigation of possibly anomalous superconducting behavior.

Such a discrepancy does not establish that a material is an unconventional superconductor. It may reflect limitations of the theoretical model, specifics of experimental samples, structural phase, stoichiometry, or source-data error. The selected materials are therefore candidates for additional physical analysis, not confirmed unconventional superconductors. The practical value of the experiment is that the pipeline can assemble a literature-derived dataset from a large publication corpus, after which an expert can compare it with independent theoretical data and inspect deviations of interest.

\subsection{Limitations}
\label{sec:discussion:limits}

The main limitations are specialization to a single domain, the modest size of the manually verified gold set, dependence of downstream results on PDF processing and extraction quality, noisy extracted metadata in hybrid field retrieval, weak intra-document evidence ranking relative to document recall, small subsets behind some per-type and conditional metrics, and dependence of the screening application on the theoretical reference dataset. There is also a trade-off between conservative $T_c$ extraction, which leaves more empty fields, and aggressive extraction, which increases false positives.

\section{Conclusion}
\label{sec:conclusion}

This paper presented a domain-aware pipeline for automatic analysis of superconductivity literature. The system combines construction of a thematic arXiv corpus, processing of scientific PDF files, extraction of materials and critical temperatures $T_c$, structured metadata, hybrid retrieval, and retrieval-augmented generation.

On the primary positive extraction set, domain-aware methods raise material--$T_c$ pair accuracy from $9.1\%$ for regex to $45.5\%$ for the full extractor, although current filters still trade recall for precision. On a 100-question domain benchmark, BM25 is strongest for formula-centered $T_c$ lookup, dense retrieval helps open questions, and hybrid retrieval helps multi-document comparison and threshold queries. Hybrid retrieval with extracted fields does not improve aggregate source recall over plain hybrid; it gives the best answer accuracy when gold evidence is already in context, but on small subsets. Qualitative cases indicate that failures often arise from evidence-span selection inside an otherwise correct paper rather than from unconstrained language-model hallucination.

Practical applicability was further demonstrated by matching the automatically assembled literature-derived dataset to an independent theoretical reference set. After expert normalization and verification, 200 materials appeared in both sources. For 20 of them the median experimental $T_c$ from the literature exceeded the theoretical reference value, which identified them as candidates for further physical study. These candidates are not interpreted as proven unconventional superconductors; the discrepancy serves only as a selection criterion.

Future work includes tightening temperature filters without losing correct candidates, more precise intra-document evidence ranking, a larger gold set and benchmark, additional physical parameters, and application of the same approach to other materials-informatics tasks.

\appendices

\section{Annotated Dataset Description}
\label{app:dataset}

The gold set contains 59 manually verified superconducting materials, of which 46 have confirmed experimental $T_c$ and 33 have an arXiv example of correct automatic extraction. The record schema uses the fields
\texttt{arxiv\_id}, \texttt{material\_gold}, \texttt{tc\_gold},
\texttt{link\_ok}, and \texttt{notes}.
The unit of evaluation is a material record. Primary material excludes comparison-only and review-only mentions. Accepted $T_c$ excludes Debye, structural, magnetic, and synthesis temperatures. A numeric match is insufficient unless the temperature belongs to the annotated material and describes the superconducting transition. Alias and formula variants are normalized before matching. Incorrect extractions retained in the gold set are used for error analysis.

\section*{Acknowledgment}

The authors thank collaborators of the ``AI Co-Researcher in Quantum Materials'' project funded within  Priority 2030
Federal Academic Leadership Program  for dataset annotation and infrastructure support.

During the preparation of this manuscript, the authors used Claude Opus 5.5 (Anthropic) for language editing and for restructuring and condensing the text in all sections. The tool was not used to generate data, run experiments, produce results, or draw scientific conclusions. All AI-assisted text was reviewed and revised by the authors, who take full responsibility for the content of the publication.

\begin{biography}
[{\includegraphics[width=1in,height=1.25in,clip,keepaspectratio]{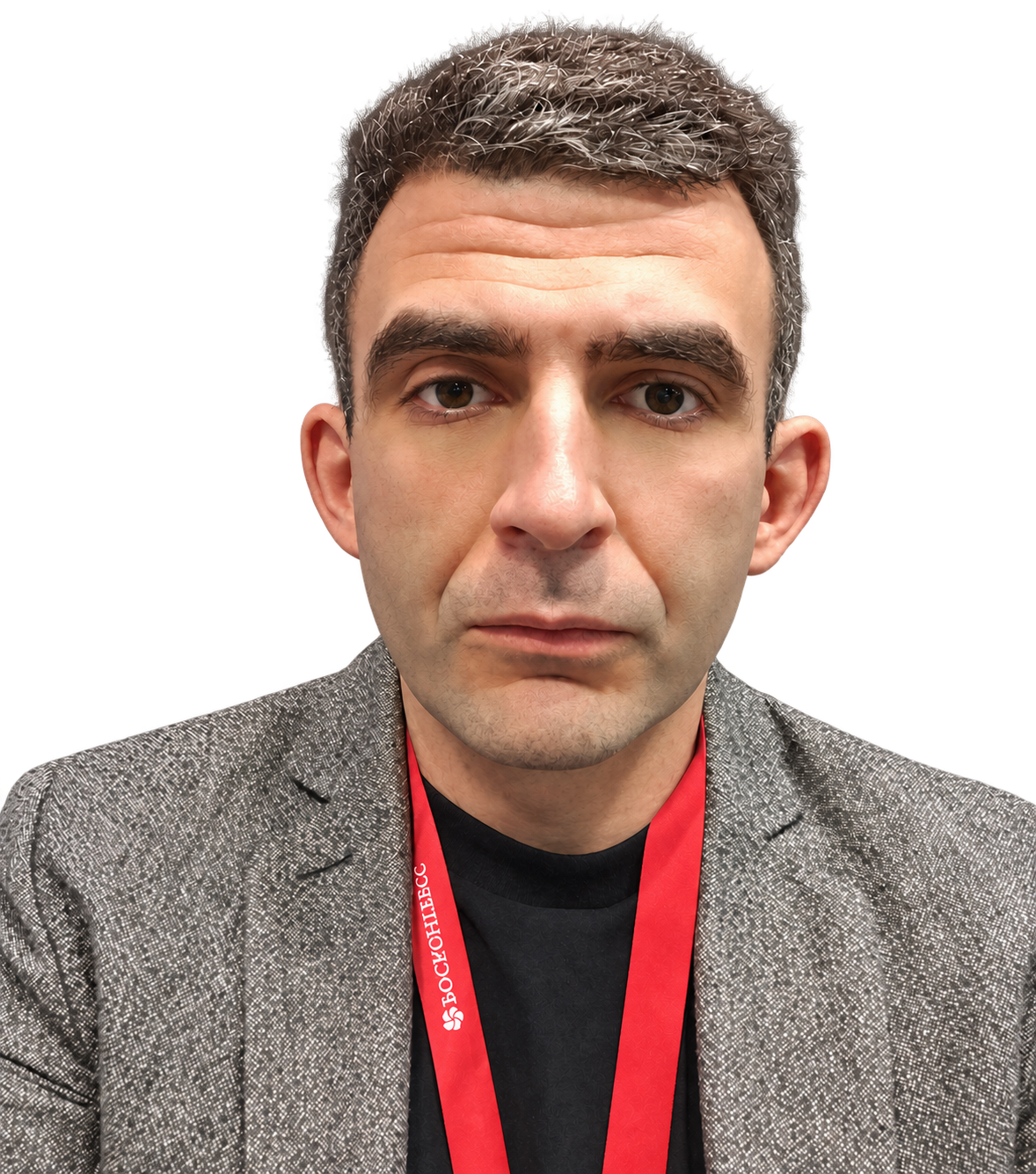}}]{Ivan Iorsh}
received the B.Sc.\ degree in physics from Saint Petersburg State Polytechnic University, Saint Petersburg, Russia, and the M.Sc.\ degree in physics from the Academic University of the Russian Academy of Sciences in 2009. He obtained the Ph.D.\ degree in condensed matter physics from Durham University, Durham, U.K., in 2013, and later defended the Doctor of Science dissertation at ITMO University, Saint Petersburg, Russia, in 2018. He has been working at ITMO University since 2013, with main specialization in condensed matter theory and quantum optics. He developed the theory describing quantum nonlinear optical processes in metamaterials and pioneered the studies of resonant optical nonlinearities in low-dimensional van der Waals materials.
\end{biography}

\begin{biography}[{\includegraphics[width=1in,height=1.25in,clip,keepaspectratio]{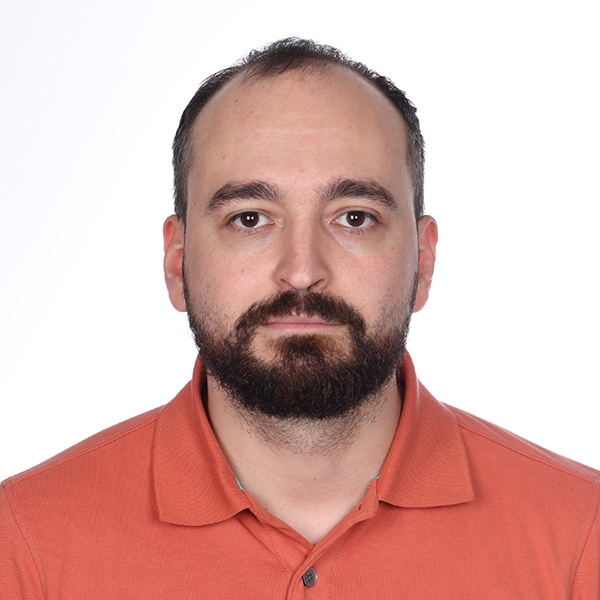}}]{Aleksandr Kapitonov}
received the B.S.\ degree in automation and control in 2009 and the M.S.\ degree in automation and control in 2011 from ITMO University, Saint Petersburg, Russia. He earned the Ph.D.\ degree in industrial automation from ITMO University in 2014. He was an Associate Professor and Dean of the Faculty of ICT at ITMO University, where he also participated in the Adaptive and Nonlinear Control Systems Laboratory. He is currently affiliated with New Uzbekistan University, Tashkent, Uzbekistan. His research interests include robot economics, blockchain technology applications in robotics, autonomous systems and UAVs, control theory, multi-agent systems, and the integration of large language model (LLM) agents in robotic systems. He has pioneered work on blockchain-based protocols for autonomous business activities in multi-agent systems and the development of smart factory concepts using decentralized technologies.
\end{biography}

\begin{biography}[{\includegraphics[width=1in,height=1.25in,clip,keepaspectratio]{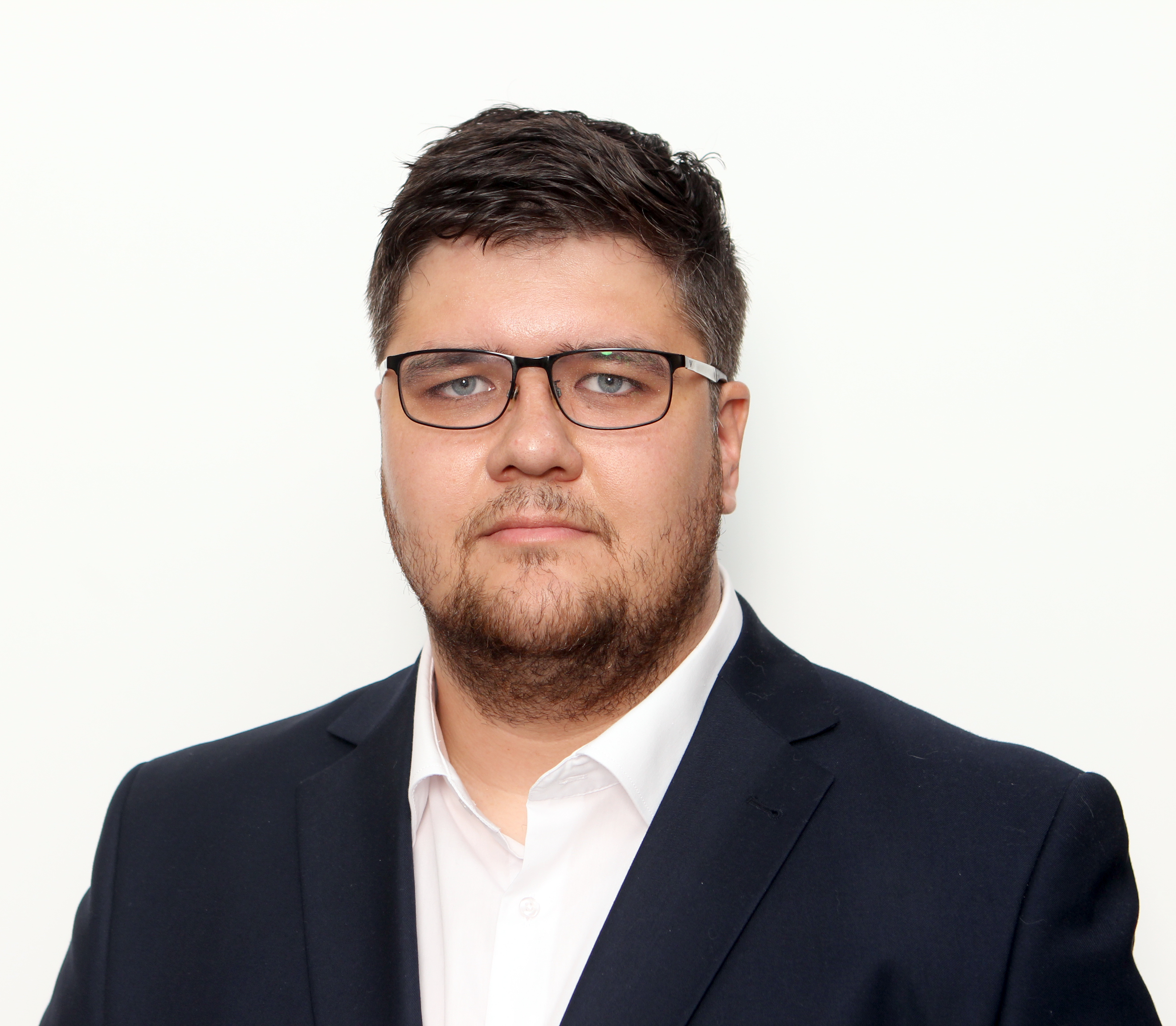}}]{Ivan Filianin}
received the B.S.\ degree in infocommunication technologies and communication systems from Pacific National University (PNU), Khabarovsk, Russia, in 2020, and the M.S.\ degree in the same specialty from ITMO University, Saint Petersburg, Russia, in 2022. He completed the Ph.D.\ program in systems, networks and devices of telecommunications (specialty 2.2.15) at ITMO University in 2026 and is preparing for the thesis defense. He is currently a researcher at New Uzbekistan University, Tashkent, Uzbekistan. His research interests include multi-access edge computing (MEC), communication networks, and machine learning applications in network operations.
\end{biography}

\begin{biography}[{\includegraphics[width=1in,height=1.25in,clip,keepaspectratio]{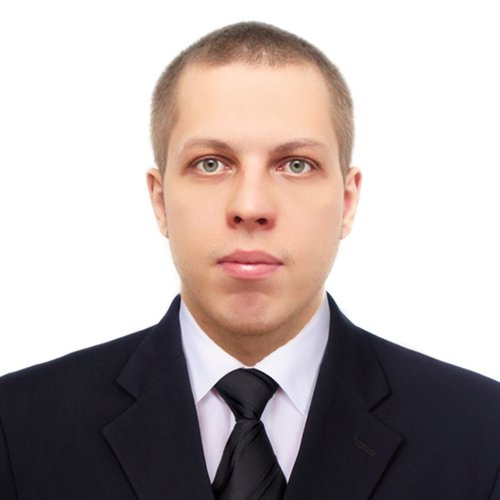}}]{Matvey Lesonen}
received the B.S.\ degree from ITMO University, Saint Petersburg, Russia, in the Department of Computational Technologies, Faculty of Information Systems Programming. He received the M.S.\ degree from ITMO University in the Department of Software Engineering and Computer Systems.
\end{biography}

\begin{biography}[{\includegraphics[width=1in,height=1.25in,clip,keepaspectratio]{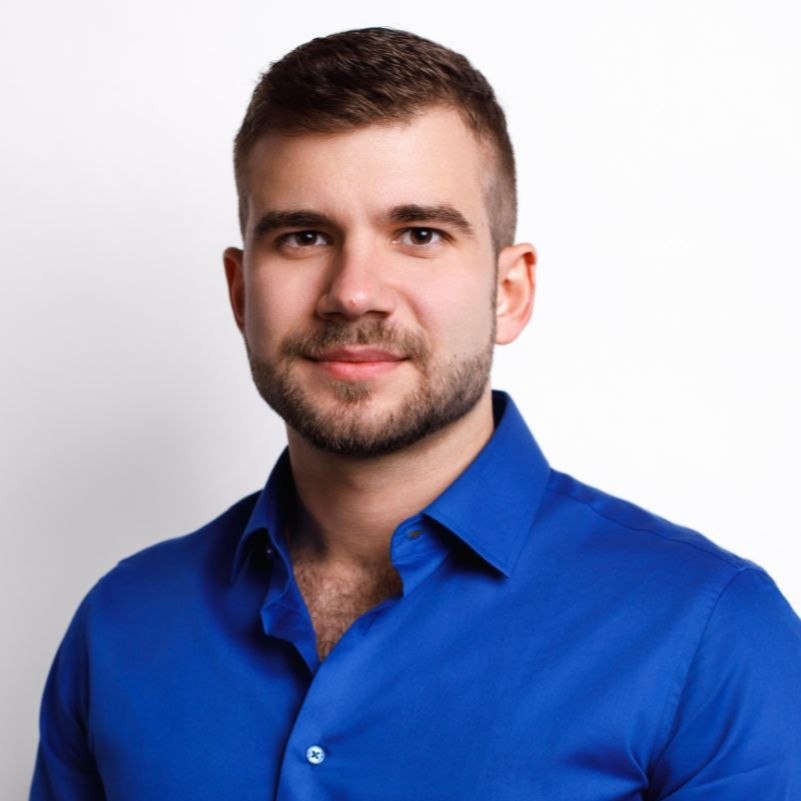}}]{Nikita Chizhov}
received the B.S.\ degree in software engineering from Peter the Great St.\ Petersburg Polytechnic University (SPbPU), Saint Petersburg, Russia, at the Institute of Computer Science and Cybersecurity. He received the M.S.\ degree in infocommunication technologies and communication systems from ITMO University, Saint Petersburg, Russia, at the Faculty of Applied Informatics.
\end{biography}

\EOD

\begin{thebibliography}{00}
\bibitem{b1} M. C. Swain and J. M. Cole, ``ChemDataExtractor: A toolkit for automated extraction of chemical information from the scientific literature,'' \emph{J. Chem. Inf. Model.}, vol. 56, no. 10, pp. 1894--1904, 2016.
\bibitem{b2} J. Mavračić \emph{et al.}, ``ChemDataExtractor 2.0: Autopopulated ontologies for materials science,'' \emph{J. Chem. Inf. Model.}, vol. 61, no. 9, pp. 4280--4289, 2021.
\bibitem{b3} I. Beltagy, K. Lo, and A. Cohan, ``SciBERT: A pretrained language model for scientific text,'' in \emph{Proc. EMNLP-IJCNLP}, 2019.
\bibitem{b4} T. Gupta \emph{et al.}, ``MatSciBERT: A materials domain language model for text mining and information extraction,'' \emph{npj Comput. Mater.}, vol. 8, Art. no. 102, 2022.
\bibitem{b5} K. Yamaguchi, R. Asahi, and Y. Sasaki, ``SC-CoMIcs: A superconductivity corpus for materials informatics,'' \emph{Sci. Technol. Adv. Mater.}, vol. 21, no. 1, pp. 235--245, 2020.
\bibitem{b6} L. Foppiano \emph{et al.}, ``SuperMat: Construction of a linked annotated dataset from superconductors-related publications,'' \emph{Sci. Data}, vol. 8, Art. no. 230, 2021.
\bibitem{b7} L. Foppiano \emph{et al.}, ``Automatic extraction of materials and properties from superconductors scientific literature,'' \emph{Sci. Rep.}, vol. 12, Art. no. 9866, 2022.
\bibitem{b8} K. Lo \emph{et al.}, ``S2ORC: The semantic scholar open research corpus,'' in \emph{Proc. ACL}, 2020.
\bibitem{b9} V. Karpukhin \emph{et al.}, ``Dense passage retrieval for open-domain question answering,'' in \emph{Proc. EMNLP}, 2020.
\bibitem{b10} A. Cohan \emph{et al.}, ``SPECTER: Document-level representation learning using citation-informed transformers,'' in \emph{Proc. ACL}, 2020.
\bibitem{b11} M. Ostendorff \emph{et al.}, ``Neighborhood contrastive learning for scientific document representations with citation embeddings,'' in \emph{Proc. EMNLP}, 2022.
\bibitem{b12} N. Thakur \emph{et al.}, ``BEIR: A heterogeneous benchmark for zero-shot evaluation of information retrieval models,'' in \emph{Proc. NeurIPS Datasets and Benchmarks}, 2021.
\bibitem{b13} P. Mandikal and R. Mooney, ``Sparse meets dense: A hybrid approach to enhance scientific document retrieval,'' 2024.
\bibitem{b14} P. Lewis \emph{et al.}, ``Retrieval-augmented generation for knowledge-intensive NLP tasks,'' in \emph{Proc. NeurIPS}, 2020.
\bibitem{b15} P. Dasigi \emph{et al.}, ``A dataset of information-seeking questions and answers anchored in research papers,'' in \emph{Proc. NAACL}, 2021.
\bibitem{b16} J. L\'ala \emph{et al.}, ``PaperQA: Retrieval-augmented generative agent for scientific research,'' 2023.
\end{thebibliography}
\end{document}